\documentclass[11pt]{article}

\input{epsf.tex}

\renewcommand{\appendix}[1]{{\Large \bf Ap\'endice: {#1}} \par}

\newcommand{\cl}[1]{{\cal{#1}}}

\renewcommand{\d}{{\rm d}}
\newcommand{\I}{{\rm i}}

\newcommand{\dfrac}[2]{\displaystyle\frac{#1}{#2}}

\newcommand{\dero}[2]{\dfrac{\d{#1}}{\d{#2}}}

\newcommand{\be}{\begin{equation}}
\newcommand{\ee}{\end{equation}}
\newcommand{\bea}{\begin{eqnarray}}
\newcommand{\eea}{\end{eqnarray}}
\newcommand{\bas}{\begin{eqnarray*}}
\newcommand{\eas}{\end{eqnarray*}}
\newcommand{\ba}{\begin{array}}
\newcommand{\ea}{\end{array}}
\newcommand{\bedes}{\begin{description}}
\newcommand{\edes}{\end{description}}

\title{{\bf Hawking Radiation in a Pleba\'nski-Demia\'nski Black Hole}}
\author{Jes\'us A.~C\'azares\footnote{cazares@irb.hr}\\
   {}\\
   {\it Rudjer Bo\v skovi\'c Institute,}\\
   {\it P.O.Box 180, HR-10002 Zagreb, Croatia}}
\date{\today}
\begin{document}

\maketitle

 \abstract{In this paper, we show the flux of Hawking radiation in a Pleba\'nski-Demia\'nski black hole from the point of view of gauge and gravitational anomalies.
 We will use the consistent anomaly method to guarantee that our results are valid in the de Sitter space.
 This is because we are including the cosmological constant into our parameters and the covariant anomaly method gives a wrong value for the Hawking temperature.
 We also show that these calculations are a general result. 
 In order to verify the consistence of our results, we can reproduce earlier known results as certain limiting cases.}

\section{Introduction}

  Black hole radiation, also called Hawking radiation, was originally reported by Zeldovich and Starobinsky in 1971 \cite{ZS-71} and it is one of the more interesting known effects.
  This effect is a consequence of the combination of two modern theories: quantum field theory and general relativity. The Hawking radiation originates by the quantization of matter in a background space--time with an event horizon, like a black hole.
  It has been found that the occupation number spectrum of quantum field modes in the vacuum state corresponds to the blackbody at a fixed temperature given by the surface gravity of the horizon.
  
  There are several explanations for Hawking radiation. The derivation by Hawking \cite{Haw-74, Haw-75} is direct and physical. He calculates the Bogoliubov coefficients between the in and out states of fields in the black hole background.
  The derivation based in quantum gravity \cite{GH-77-1, GH-77-2} is fast and elegant, but needs microscopic foundation.
  The derivation based on string theory \cite{SV-96, Peet-00} has a logical consistency foundation but can be applied only to special solutions and does not explain the simplicity and generality of the results inferred from other methods.
  
  Christensen and Fulling \cite{CF-77} showed that the magnitude of the Hawking blackbody effect at infinity is directly proportional to the magnitude of the trace anomaly. With this starting point, Robinson and Wilczek \cite{RW-05} have recently shown a procedure for calculating Hawking radiation via gravitational anomaly cancelation in a Schwarzschild black hole metric.
  This original idea soon was extended to the Reissner-N\"ordstrom metric \cite{IUW-06-1}, to rotating black holes \cite{IUW-06-2, MS-06, XC-07, JW-07, IMU-06} and even to metrics with NUT parameter \cite{LCY-08}. This leads to the important question: What are the fluxes like for Hawking radiation in the most general black hole?

  As it is known, the most general of Petrov D--type solutions of the Einstein-Maxwell equations is the so-called Pleba\'nski-Demia\'nski solution \cite{PD-76}. In the present article, we calculate the Hawking radiation for the Pleba\'nski-Demia\'nski black hole in the spirit of Robinson and Wilczek method and show the form of the fluxes and the Hawking temperature.
  It is necessary to comment that the method based on the gravitational and gauge anomalies reproduce succesfully the Hawking fluxes, but this anomalies contain only the information of fluxes of energy and charge.
  These fluxes correspond to the zero-th and first moments of the thermal distribution of radiation. To obtain the full information, it is necessary to calculate all the other higher moments. Iso, Morita and Umetsu \cite{IMU-07-1, IMU-07-2, IMU-08-1, IMU-08-2} attributed these higher fluxes to phenomenological higher spin currents, {\it i.e.} higher spin generalizations of the energy-momentum tensor.
  Bonora, Cvitan, Pallua and Smoli\'c \cite{BC-08, BCPS-08, BCPS-09} have shown that such higher currents describe the higher spin fluxes of the Hawking radiation and that these higher spin currents cannot have trace anomalies neither have diffeomorphism anomalies. They also showed that the thermal spectrum of the Hawking radiation is induced by the underlying $W_{1+\infty}$ algebra structure of the higher spin currents.
  That is, the Hawking radiation and in particular its thermal spectrum, points toward the existence of a symmetry much larger than the Virasoro algebra in the near horizon region. That is a $W_\infty$ or a $W_{1+\infty}$ algebra, an extension of the $W_\infty$ algebra to include a $U(1)$ current.

  The article is organized as follows: In the second section we introduce the Pleba\'nski-Demia\'nski metric and by a partial wave decomposition of the scalar field in terms of spherical harmonics, we get the effective action, corresponding to a $(1+1)$--dimensional metric in a dilaton background and a gauge field. In the present situation, we will get three different gauge fields: one related to the mass and the other two are related to the electric and magnetic charges, respectively.
  In sections 3 and 4, using the generalization made by Vagenas and Das \cite{VD-06, DRV-07}, we calculate the gauge and gravitational anomalies from which the Hawking radiation arises at the event horizon and cosmological horizon, respectively. It is necessary to comment at this point that we shall use the consistent anomaly method because we are including the cosmological constant into our parameters, since the covariant anomaly method gives a wrong value for the Hawking temperature \cite{APGS-08}.
  Recently, a new method has been found by Zampeli, Singleton and Vagenas \cite{ZSV-12} that works perfectly for de Sitter and Rindler space--time using either the covariant or the consistent formalism.
  In order to verify that our results have mathematical and physical consistence, we reduce several parameters of our metric in section 5 to get two well-known metrics which in the limiting cases correspond to the Kerr-Newman-de Sitter black hole metric and the NUT-Kerr-Newman-de Sitter. When this parameters are also reduced in the obtained fluxes, we recover the same results obtained previously in the literature \cite{JW-07, LCY-08}.
  From this two metrics, we can vanish all the parameters except the cosmological constant and get the de Sitter metric. When we do this, we  get the right Hawking temperature \cite{Ji-07, APGS-08}.
  Finally, the conclusion is presented in section 6.


\section{The Effective action of Pleba\'nski-Demia\'nski black hole}

  In the Petrov clasification, the most general D--type metric is the Pleba\'nski-Demia\'nski metric \cite{GP-06}. It is possible to get, by certain transformations and limiting procedures, many particular and well known D--type metrics, such as the Pleba\'nski-Carter, the Kerr-Newman, the Kerr and Schwarzschild solutions among others. The Pleba\'nski-Demia\'nski metric is \cite{PD-76}
     \be
     \d s^2=\dfrac1{(1-\hat p\hat r)^2}\left[\dfrac{{\cl Q}(\d\hat\tau-\hat p^2\d\hat\sigma)^2}{\hat r^2+\hat p^2}-\dfrac{{\cl P}(\d\hat\tau+\hat r^2\d\hat\sigma)^2}{\hat r^2+\hat p^2}-(\hat r^2+\hat p^2)\Bigl(\dfrac{\d\hat p^2}{\cl P}+\dfrac{\d\hat r^2}{\cl Q}\Bigr)\right],
     \ee
  with the 1-form potential
     \be
     \cl A=\dfrac{\hat e+\I\hat g}{\hat r+\I\hat p}\Bigl(\d\hat\tau-\I\hat p\hat r\d\hat\sigma\Bigr);
     \ee
  the functions $\cl P$ and $\cl Q$ in the metric, are the related quartic functions
     \bea
     \bigskip \cl P & = & \hat k+2\hat n\hat p-\hat\epsilon\hat p^2+2\hat m\hat p^3-\left(\hat k+\hat e^2+\hat g^2+\frac\Lambda3\right)\hat p^4, \\
     \cl Q & = & (\hat k+\hat e^2+\hat g^2)-2\hat m\hat r+\hat\epsilon\hat r^2-2\hat n\hat r^3-\left(\hat k+\frac\Lambda3\right)\hat r^4,
     \eea
  and $\hat m,\,\hat n,\,\hat e,\,\hat g,\,\hat\epsilon,\,\hat k,\,\Lambda$ are arbitrary real parameters. It is habitually assumed that $\hat m$ and $\hat n$ are the mass and NUT parameters, respectively, but this is not generally the case. To describe the complete family of black hole--like space--time, recently Griffiths and Podolsky \cite{GP-06} have shown that it is possible to transform the Pleba\'nski-Demia\'nski metric in the following form
     \be
     \d s^2=\dfrac1{\Omega^2}\left\{\dfrac{\hat\cl Q}{\rho^2}\left[\d t-\dfrac{a\sin^2\theta+4l\sin^2\dfrac\theta2}\Xi\d\varphi\right]^2-\dfrac{\rho^2}{\hat\cl Q}\d r^2-\dfrac{\hat\cl P}{\rho^2}\left[a\d t-\dfrac{r^2+(a+l)^2}\Xi\d\varphi\right]^2-\dfrac{\rho^2}{\hat\cl P}\sin^2\theta\d\theta^2\right\},
     \label{PD}
     \ee
  and, using the same transformations given in \cite{GP-06}, the potential becomes
     \bea
     \bigskip \cl A & = & A_\mu+\I B_\mu \\
     \bigskip & = & \left(\dfrac{Qr+G(l+a\cos\theta)}{r^2+(l+a\cos\theta)^2},0,0,-\dfrac{Qr\Bigl(4al\sin^2\dfrac\theta2+a^2\sin^2\theta\Bigr)+G(l+a\cos\theta)\Bigl(r^2+(l+a)^2\Bigr)}{a\Xi\Bigl(r^2+(l+a\cos\theta)^2\Bigr)}\right) \nonumber \\
     & & +\I\left(\dfrac{Gr-Q(l+a\cos\theta)}{r^2+(l+a\cos\theta)^2},0,0,\dfrac{-Gr\Bigl(4al\sin^2\dfrac\theta2+a^2\sin^2\theta\Bigr)+Q(l+a\cos\theta)\Bigl(r^2+(l+a)^2\Bigr)}{a\Xi\Bigl(r^2+(l+a\cos\theta)^2\Bigr)}\right); \label{pot-PD}
     \eea
  where $A_\mu$ is the electric potential, $B_\mu$ is a magnetic--like potential, and the functions that appear in the metric are given by
     \bea
     \Omega & = & 1-\dfrac\alpha\omega(l+a\cos\theta)r, \label{funcmetric-1} \\
     \rho^2 & = & r^2+(l+a\cos\theta)^2, \\
     \Xi & = & 1+\dfrac\Lambda3(a^2-l^2), \\
     {\hat\cl P} & = & \sin^2\theta(1-a_3\cos\theta-a_4\cos^2\theta), \\
     {\hat\cl Q} & = & (\omega^2k+Q^2+G^2)-2mr+\epsilon r^2-2\alpha\dfrac n\omega r^3-\left(\alpha^2k+\dfrac\Lambda3\right)r^4, \label{funcmetric-5} \\
     a_3 & = & 2\alpha\dfrac a\omega m-4\alpha^2\dfrac{al}{\omega^2}(\omega^2k+Q^2+G^2)-4\dfrac\Lambda3al, \label{funcmetric-a3} \\
     a_4 & = & -\alpha^2\dfrac{a^2}{\omega^2}(\omega^2k+Q^2+G^2)-\dfrac\Lambda3a^2, \label{funcmetric-a4} \\
     \epsilon & = & \dfrac{\omega^2k}{a^2-l^2}+4\alpha\dfrac l\omega m-(a^2+3l^2)\left(\dfrac{\alpha^2}{\omega^2}(\omega^2k+Q^2+G^2)+\dfrac\Lambda3\right), \label{funcmetric-e} \\
     n & = & \dfrac{\omega^2kl}{a^2-l^2}-\alpha\dfrac{a^2+l^2}\omega m+(a^2-l^2)l\left(\dfrac{\alpha^2}{\omega^2}(\omega^2k+Q^2+G^2)+\dfrac\Lambda3\right), \label{funcmetric-n} \\
     \left(\dfrac{\omega^2}{a^2-l^2}+3\alpha^2l^2\right)k & = & 1+2\alpha\dfrac l\omega m-3\alpha^2\dfrac{l^2}{\omega^2}(Q^2+G^2)-l^2\Lambda. \label{constriction}
     \eea

  Now $m,\,Q,\,G,\,a,\,l,\,\alpha,\,\Lambda$ represent the mass, electric and magnetic charge, angular momentum, NUT parameter, acceleration and cosmological constant, respectively. That is, they are now physical parameters.
  It is assumed that $\vert a_3\vert$ and $\vert a_4\vert$ are sufficiently small to guarantee that $\hat\cl P$ has no additional roots in $\theta\in[0,\pi]$. The equations (\ref{funcmetric-e}--\ref{constriction}) define the parameters $\epsilon,\,n,\,k$ and give a strong restriction to $\omega$. It can take a convenient value if $a$ and $l$ are not both zero.
  For simplicity in our calculations, it is possible to write the metric (\ref{PD}) in the following form
     \be
     \d s^2=\dfrac1{\Omega^2}\left\{\dfrac{{\hat\cl Q}-a^2{\hat\cl P}}{\rho^2}\d t^2-\dfrac2{\rho^2\Xi}\Bigl({\hat \cl Q}\Theta-a{\hat\cl P}R\Bigr)\d t\d\varphi+\dfrac1{\rho^2\Xi^2}\Bigl({\hat\cl Q}\Theta^2-{\hat\cl P}R^2\Bigr)\d\varphi^2-\dfrac{\rho^2}{\hat\cl Q}\d r^2-\dfrac{\rho^2\sin^2\theta}{\hat\cl P}\d\theta^2\right\},
     \label{PD2}
     \ee
  where we have introduced the functions $\Theta=\Theta(\theta)$ and $R=R(r)$ given by
     \be
     \Theta=a\sin^2\theta+4l\sin^2\theta/2\qquad,\qquad\qquad\qquad R=r^2+(a+l)^2.
     \ee

  The outer horizon $(r=r_H)$ is determined by $\hat\cl Q(r_H)=0$. The term $1/\Omega^2$ is a conformal factor, so, this term does not contribute to the Hawking radiation, and we can omit it in our calculations. We will consider matter fields in the Pleba\'nski-Demi\'anski black hole background. As we have both electric and magnetic charge, we can take into consideration that the covariant derivative is \cite{Zwa-71, Sin-95}
     \be
     D_\mu=\partial_\mu-\I eA_\mu-\I gB_\mu,
     \label{derivadacovariante}
     \ee
  where $e$ and $g$ are the electric and magnetic charges, respectively. With all this taken into in consideration, the action is given by
     \bea
     S & = & \frac12\int\d^4x\sqrt{-g}g^{\mu\nu}D_\mu\Phi D_\nu\Phi \\
     & = & -\frac12\int\d^4x\sqrt{-g}\Phi^*\left\{g^{tt}\Bigl[\partial_t^2-2\I(eA_t+gB_t)\partial_t-(eA_t+gB_t)^2\Bigr]+2g^{t\varphi}\Bigl[\partial_t\partial_\varphi-\I(eA_\varphi+gB_\varphi)\partial_t\right. \nonumber \\
     & & \qquad-\I(eA_t+gB_t)\partial_\varphi-(eA_t+gB_t)(eA_\varphi+gB_\varphi)\Bigr]+g^{\varphi\varphi}\Bigl[\partial_\varphi^2-2\I(eA_\varphi+gB_\varphi)\partial_\varphi-(eA_\varphi+gB_\varphi)^2\Bigr] \nonumber \\
     & & \qquad+\left.\dfrac1{\sqrt{-g}}\partial_r(\sqrt{-g}g^{rr}\partial_r)+\dfrac1{\sqrt{-g}}\partial_\theta(\sqrt{-g}g^{\theta\theta}\partial_\theta)\right\}\Phi, \label{action}
     \eea
  where the elements of the metric are
     \be
     \ba{lll}
     \bigskip g^{tt}=-\dfrac{\rho^2({\hat\cl Q}\Theta^2-R^2{\hat\cl P})}{{\hat\cl P}{\hat\cl Q}(R-a\Theta)^2},\qquad\qquad & g^{t\varphi}=-\dfrac{\rho^2\Xi({\hat\cl Q}\Theta-aR{\hat\cl P})}{{\hat\cl P}{\hat\cl Q}(R-a\Theta)^2},\qquad\qquad & g^{\varphi\varphi}=-\dfrac{\rho^2\Xi^2({\hat\cl Q}-a^2{\hat\cl P})}{{\hat\cl P}{\hat\cl Q}(R-a\Theta)^2}, \\
     g^{rr}=-\dfrac{\hat\cl Q}{\rho^2}, & g^{\theta\theta}=-\dfrac{\hat\cl P}{\rho^2\sin^2\theta}, & \sqrt{-g}=\dfrac{(R-a\Theta)\sin\theta}\Xi.
     \ea\ee

  Performing the partial wave decomposition of the scalar field $\Phi$ in terms of the spherical harmonics $\Phi=\sum\limits_{l,m}\phi_{lm}(r,t)Y_{lm}(\theta,\varphi)$ this action can be reduced to a two dimensional effective theory. To do this it is necessary to transform it to the $r_*$ tortoise coordinate defined by
     \be
     \dero{r_*}r=\dfrac R{\hat\cl Q}=f(r)^{-1},
     \label{f-tortuga}
     \ee
  and considering the region near the horizon. After this process, the action (\ref{action}) can be simplified to
     \be
     S=-\dfrac12\int\d t\d r\dfrac R\Xi\phi_{l,m}^*\left\{\dfrac R{\hat\cl Q}\left(\partial_t+\I m\dfrac{a\Xi}R+\I eA_S+\I g\breve A_S\right)^2-\partial_r\dfrac{\hat\cl Q}R\partial_r\right\}\phi_{l,m};
     \label{actionefectiva}
     \ee
  that is, each partial wave of the scalar field can be considered as a $(1+1)$--dimensional complex scalar field in the background of the dilaton $\Psi$, where the elements of this metric $\tilde g_{\mu\nu}$ and gauge fields $\tilde A_\mu$ are given by
     \bea
     \Psi & = & \dfrac R\Xi=\dfrac{r^2+(a+l)^2}{1+\dfrac\Lambda3(a^2-l^2)}, \\
     \tilde g_{tt} & = & -\dfrac{\hat\cl Q}R=-\dfrac{(\omega^2k+Q^2+G^2)-2mr+\epsilon r^2-2\alpha\dfrac n\omega r^3-\left(\alpha^2k+\dfrac\Lambda3\right)r^4}{r^2+(a+l)^2}, \\
     \tilde g_{rr} & = & \dfrac R{\hat\cl Q}, \\
     \tilde A_{t1} & = & -\dfrac{a\Xi}R=-\dfrac{a\Bigl(1+\dfrac\Lambda3(a^2-l^2)\Bigr)}{r^2+(a+l)^2}, \label{gauge-t1} \\
     \tilde A_{t2} & = & -A_S=-\dfrac{Qr-G\dfrac\Lambda3(l+a)(a^2-l^2)}{r^2+(a+l)^2}, \label{gauge-t2} \\
     \tilde A_{t3} & = & -\breve A_S=-\dfrac{Gr+Q\dfrac\Lambda3(l+a)(a^2-l^2)}{r^2+(a+l)^2}, \label{gauge-t3} \\
     \tilde A_r & = & 0;
     \eea
  where $m,\,e,\,g$ are the charges of the gauge fields $\tilde A_{t1},\,\tilde A_{t2},\,\tilde A_{t3}$; respectively.

  In this $(1+1)$--dimensional reduction, the effective field theory is based on the observable physics and defined outside the horizon of the black hole. This means that the ingoing modes are omitted at the horizon making the theory chiral there. With this, each partial wave becomes anomalous with respect to gauge and general coordinate symmetries.
  So, in order to have gauge invariance and diffeomorphism covariance it is necessary that the fluxes of the $U(1)$ gauge current and the energy-momentum tensor cancel the gauge and gravitational anomaly at the horizon, respectively. We will show this procedure in the next section.


\section{Anomalies and Hawking radiation}


  An anomaly in QFT is a conflict between a symmetry from the classical action and the quantization. There exist anomalies in global symmetries and gauge symmetries  \cite{Ber-96, Bil-08}.
  The gauge current must satisfy the conservation equation $\nabla_\mu J^\mu=0$. However, near the horizon, the $U(1)$ gauge current satisfies an anomalous equation \cite{Ber-96}
     \be
     \nabla_\mu J^\mu=\dfrac\alpha{4\pi\sqrt{-g}}\epsilon^{\mu\nu}\partial_\mu A_\nu,
     \ee
  where we have used $\alpha$ to denote the gauge charge of the $U(1)$ gauge field $A_\nu$. So, in the region $r\geq r_H+\epsilon$, as there is no anomaly, the $U(1)$ gauge current must satisfy the conservation equation $\partial_r J^r_{(out)}=0$ but, near the horizon it must satisfy $\partial_r J^r_{(H)}=\dfrac{\alpha\partial_r\tilde A_t}{4\pi}$. Thus we can get
     \bea
     J^r_{(out)} & = & c_o, \qquad\qquad\qquad\qquad\qquad\qquad\qquad\,\,\,\,\, r\geq r_H+\epsilon, \label{currentout} \\
     J^r_{(H)} & = & c_H+\dfrac1{4\pi}\Bigl(\tilde A_t(r)-\tilde A_t(r_H)\Bigr),\qquad\qquad r_H\leq r\leq r_H+\epsilon; \label{currenth}
     \eea
  where $c_o$ and $c_H$ are integration constants. $c_o$ is the value of the current at $r=\infty$ and $c_H$ is the value of the consistent current of the outgoing modes at the horizon. To get the values of $c_o$ and $c_H$, we can use the consistent current
     \be
     J^\mu=J^\mu_{(out)}\Theta(r-r_H-\epsilon)+J^\mu_{(H)}\left[1-\Theta(r-r_H-\epsilon)\right],
     \ee
  where the scalar step function is
     \be
     \Theta(r-r_H-\epsilon)=\left\{\ba{ll}
     \bigskip 1,\qquad & r\geq r_H+\epsilon, \\
     0, & r_H\leq r\leq r_H+\epsilon.
     \ea\right.
     \ee

  Since we have omitted the ingoing modes near the horizon, this current is only a part of the total current.

  If a classical action $S[\Phi,g_{\mu\nu}]$ is quantized, we get
     \be
     \cl W[g_{\mu\nu}]=-\I\ln\left(\int\cl D\Phi e^{\I S[\Phi,g_{\mu\nu}]}\right);
     \label{acFeynman}
     \ee
  and, under gauge transformations, the variation of the quantum effective action is
     \be
     -\delta\cl W=\int\d t\d r\sqrt{-\tilde g}\lambda\nabla_\mu J^\mu,
     \ee
  where $\lambda$ is a gauge parameter. By integration by parts we have
     \be
     -\delta\cl W=\int\d t\d r\lambda\left[\delta(r-r_H-\epsilon)\Bigl(J^r_{(out)}-J^r_{(H)}+\dfrac\alpha{4\pi}\tilde A_t\Bigr)+\partial_t\Bigl(\dfrac\alpha{4\pi}\tilde A_t\left[1-\Theta(r-r_H-\epsilon)\right]\Bigr)\right].
     \ee

  The total effective action must be gauge invariant. So, the last term would vanish by quantum effects of the classically irrelevant ingoing modes. The coefficient of the delta function would also cancel, and we get
     \be
     c_o=c_H-\dfrac\alpha{4\pi}\tilde A_t(r_H).
     \ee
  To ensure the regularity requirement at the horizon, the covariant current must also vanish there. Since the covariant current is $\tilde J^r=J^r+\dfrac\alpha{4\pi}\tilde A_t\left[1-\Theta(r-r_H-\epsilon)\right]$, the condition at the horizon $\tilde J^r=0$ determines the flux of the $U(1)$ gauge current as
     \be
     c_o=-\dfrac\alpha{2\pi}\tilde A_t(r_H).
     \label{fluxgaugecurrent}
     \ee

  We have three gauge charges for a Pleba\'nski-Demia\'nski black hole; thus, the $U(1)$ gauge charge flux, the electric current flux and the magnetic current flux are determined by
     \bea
     f_m & = & -\dfrac m{2\pi}\tilde A_t(r_H)=\dfrac{m^2a\Xi+me\Bigl(Qr_H-G(a+l)(\Xi-1)\Bigr)+mg\Bigl(Gr_H+Q(a+l)(\Xi-1)\Bigr)}{2\pi\Bigl(r_H^2+(a+l)^2\Bigr)}, \label{fluxdem} \\
     f_e & = & -\dfrac e{2\pi}\tilde A_t(r_H)=\dfrac{mea\Xi+e^2\Bigl(Qr_H-G(a+l)(\Xi-1)\Bigr)+eg\Bigl(Gr_H+Q(a+l)(\Xi-1)\Bigr)}{2\pi\Bigl(r_H^2+(a+l)^2\Bigr)}, \label{fluxdeq} \\
     f_g & = & -\dfrac g{2\pi}\tilde A_t(r_H)=\dfrac{mga\Xi+ge\Bigl(Qr_H-G(a+l)(\Xi-1)\Bigr)+g^2\Bigl(Gr_H+Q(a+l)(\Xi-1)\Bigr)}{2\pi\Bigl(r_H^2+(a+l)^2\Bigr)}. \label{fluxdeg}
     \eea
\par
\vspace*{1ex}
\par
\begin{center}
 * * *
\end{center}


  The anomalies in global symmetries are theoretical inconsistences. For this reason their cancellation gives important restrictions \cite{Ber-96, Bil-08}.
  A gravitational anomaly is a gauge anomaly in general covariance, making non-conservative the energy-momentum tensor.
  This anomaly can only happen in theories with chiral matter coupled to gravity in a $(4k+2)$--dimentional space--time, with $k$ being an integer.
  This chiral matter can be a fermion and can also be a $2k$--form with an (anti--) autodual field. An important situation is the $(1+1)$--dimensional scalar autodual field; this field obeys
     \be
     \partial_a\phi=\epsilon_{ab}\partial^b\phi,
     \ee
  that is, it only has modes that are moving to the right, so it is chiral. The anomaly is \cite{AW-84, BK-01}
     \be
     \nabla_aT_b^a=\dfrac1{\sqrt{-g}}\partial_aN^a_b=\dfrac1{96\pi\sqrt{-g}}\epsilon^{cd}\partial_d\partial_a\Gamma^a_{bc};
     \label{anomaly}
     \ee
  that is, the energy-momentum tensor is not conserved in a curved space--time. As discussed previously, the effective field theory is defined outside the event horizon.
  In the region $r\geq r_H+\epsilon$ there is an effective background gauge potential, but without anomaly, so the energy-momentum tensor satisfies the modified conservation equation $\partial_r T^r_{t(out)}=c_o\partial_r\tilde A_t$. Near the horizon $r_H\leq r\leq r_H+\epsilon$ the energy-momentum tensor exhibits an anomaly and satisfies the Ward identity, that is $\partial_r T^r_{t(H)}=J^r_{(H)}\partial_r\tilde A_t+\tilde A_t\partial_r J^r_{(H)}+\partial_r N^r_t$, where $N^r_t=({f^\prime}^2+ff^{\prime\prime})/192\pi$ and in our situation this gives
     \be
     N^r_t=\dfrac1{192\pi}\dfrac{{\hat\cl Q}\cl M+\cl C^2}{R^4},
     \ee
  where the functions $\cl C$ and $\cl M$ are given explicitly by
     \bea
     \cl C & = & {\hat\cl Q}'R-2r{\hat\cl Q} \\
     & = & 2m\Bigl(r^2-(a+l)^2\Bigr)+2\Bigl(\epsilon(a+l)^2-\omega^2k-Q^2-G^2\Bigr)r-2\alpha\dfrac n\omega\Bigl(r^2+3(a+l)^2\Bigr)r^2 \nonumber \\
     & & \qquad-2\Bigl(\alpha^2k+\dfrac\Lambda3\Bigr)\Bigl(r^2+2(a+l)^2\Bigr)r^3, \\
     \cl M & = & \cl C'R-2R'\cl C \\
     & = & 8m(a+l)^2r-2\Bigl(\epsilon(a+l)^2-\omega^2k-Q^2-G^2\Bigr)\Bigl(3r^2-(a+l)^2\Bigr)+2\alpha\dfrac n\omega\Bigl(5(a+l)^2r^3-3(a+l)^4r\Bigr) \nonumber \\
     & & \qquad-2\Bigl(\alpha^2k+\dfrac\Lambda3\Bigr)\Bigl(r^6+3(a+l)^2r^4+6(a+l)^4r^2\Bigr).
     \eea
  We can use the energy-momentum tensor
     \be
     T_t^r=T^r_{t(out)}\Theta(r-r_H-\epsilon)+T^r_{t(H)}\left[1-\Theta(r-r_H-\epsilon)\right],
     \ee
  which combines contributions from these two regions.

  Under general coordinate transformations $x^\mu\longrightarrow x^\mu-\lambda^\mu$, the variation of the quantum effective action (\ref{acFeynman}) becomes
     \bea
     -\delta_\lambda\cl W & = & \int\d t\d r\lambda^\nu\nabla_\mu T^\mu_\nu \\
     & = & \int\d t\d r\lambda^t\left[c_o\partial_r\tilde A_t+\partial_r\Bigl(\dfrac\alpha{4\pi}\tilde A_t^2+N_t^r\Bigr)\left[1-\Theta(r-r_H-\epsilon)\right]\right. \nonumber \\
     & & \qquad\left.+\Bigl(T_{t(out)}^r-T^r_{t(H)}+\dfrac\alpha{4\pi}\tilde A_t^2+N_t^r\Bigr)\delta(r-r_H-\epsilon)\right];
     \eea
  where the relation $J^r_{(H)}=c_o+\dfrac\alpha{4\pi}\tilde A_t$ has been used. The effective action must vanish if we demand the covariance under the diffeomorphism transformation. The first term of the effective action is the classical effect of the background electric field. The second term is cancelled by the quantum effect of the classically irrelevant ingoing modes. The third one would be also cancelled, leading to the condition
     \be
     a_o=a_H+\dfrac\alpha{4\pi}\tilde A_t^2(r_H)-N_t^r(r_H),
     \ee
  where
     \bea
     a_o & = & T^r_{t(out)}-c_o\tilde A_t, \\
     a_H & = & T^r_{t(H)}-\int\limits_{r_H}^r\d r\partial_r\left[c_o\tilde A_t+\dfrac\alpha{4\pi}\tilde A_t^2+N_t^r\right],
     \eea
  are the values of the energy flow at infinity and at the horizon, respectively. To ensure the regularity requirement at the horizon, the covariant energy-momentum tensor must also vanish there. Since the energy-momentum tensor is $\tilde T^r_t=T^r_t+\dfrac1{192\pi}(ff^{\prime\prime}-2{f^\prime}^2)$, it will take the explicit form
     \be
     \tilde T^r_t=T^r_t+\dfrac1{192\pi}\dfrac{{\hat\cl Q}\cl M-2\cl C^2}{R^4}.
     \ee
 The condition at the horizon $\tilde T^r_t=0$ gives us
     \be
     a_H=2N_t^r(r_H)=\dfrac{\kappa^2}{24\pi};
     \ee
  where the surface gravity of the black hole is
     \bea
     \kappa\,\,=\,\,\frac12\partial_rf\Bigl\vert_{r=r_H} & = & \frac12\dfrac{\cl C}{R^2}\Bigr\vert_{r=r_H} \\
     & = & \dfrac{m\Bigl(r_H^2-(a+l)^2\Bigr)+r_H\Bigl(\epsilon(a+l)^2-\omega^2k-Q^2-G^2\Bigr)-\alpha\dfrac n\omega\Bigl(r_H^2+3(a+l)^2\Bigr)r_H^2}{\Bigl(r_H^2+(a+l)^2\Bigr)^2} \nonumber \\
     & & \qquad-\dfrac{\Bigl(\alpha^2k+\dfrac\Lambda3\Bigr)\Bigl(r_H^2+2(a+l)^2\Bigr)r_H^3}{\Bigl(r_H^2+(a+l)^2\Bigr)^2}.
     \eea
  Then, the flux of the energy-momentum tensor required to restore general coordinate covariance at quantum level in the effective field theory is
     \bea
     a_o & = & N_t^r(r_H)+\dfrac\alpha{4\pi}\tilde A_t^2(r_H) \\
     & = & \dfrac1{4\pi}\left(\dfrac{ma\Xi+e\Bigl(Qr_H-G(a+l)(\Xi-1)\Bigr)+g\Bigl(Gr_H+Q(a+l)(\Xi-1)\Bigr)}{r_H^2+(a+l)^2}\right)^2+\dfrac\pi{12}T_h^2, \label{flux}
     \eea
  where $T_h=\dfrac\kappa{2\pi}$ is the Hawking temperature of the black hole. As it is known, the Planck distribution at an inverse temperature $\beta$ with a chemical potential $\mu$ is
    \be
    N_{\pm}(\omega)=\dfrac1{e^{\beta(\omega\mp\mu)}-1}\qquad,\qquad N_{\pm}(\omega)=\dfrac1{e^{\beta(\omega\mp\mu)}+1},
    \ee
  for bosons and fermions respectively. In the zero temperature limit, if $\omega\mp\mu$ it is negative, the distribution become $\mp 1$ for bosons or fermions. In the bosonic case, this result leads to the effect of superradiance. But, in the fermionic case, the occupation numbers become 1 when temperature goes to 0 for these low frequency modes. This leads to zero flux of radiation even at the extremal case  \cite{IUW-06-1, IUW-06-2}.

  We will take in consideration the fermionic case. The Hawking distribution with chemical potential $\mu=m\tilde A_{t_1}+e\tilde A_{t_2}+g\tilde A_{t_3}$ of the Pleba\'nski-Demia\'nski black hole is given by
     \be
     N_{\pm m,\pm e,\pm g}=\dfrac1{exp\Bigl(\dfrac{\omega\mp m\tilde A_{t_1}\mp e\tilde A_{t_2}\mp g\tilde A_{t_3}}{T_h}\Bigr)+1}.
     \label{Planck}
     \ee

  From this distribution (eq. \ref{Planck}), the angular momentum flux (that is, the $U(1)$ gauge current flux), the electric current flux and the magnetic current flux can be obtained as
     \bea
     F_m & = & m\int\limits_0^\infty\dfrac1{2\pi}\Bigl[N_{m,e,g}(\omega)-N_{-m,-e,-g}(\omega)\Bigr]\d\omega \\
     & & \qquad=\dfrac{m^2a\Xi+me\Bigl(Qr_H-G(a+l)(\Xi-1)\Bigr)+mg\Bigl(Gr_H+Q(a+l)(\Xi-1)\Bigr)}{2\pi\Bigl(r_H^2+(a+l)^2\Bigr)}, \label{Hawkingdem}
     \eea
     \bea
     F_e & = & e\int\limits_0^\infty\dfrac1{2\pi}\Bigl[N_{m,e,g}(\omega)-N_{-m,-e,-g}(\omega)\Bigr]\d\omega \\
     & & \qquad=\dfrac{mea\Xi+e^2\Bigl(Qr_H-G(a+l)(\Xi-1)\Bigr)+eg\Bigl(Gr_H+Q(a+l)(\Xi-1)\Bigr)}{2\pi\Bigl(r_H^2+(a+l)^2\Bigr)}, \label{Hawkingdeq} \\
     F_g & = & g\int\limits_0^\infty\dfrac1{2\pi}\Bigl[N_{m,e,g}(\omega)-N_{-m,-e,-g}(\omega)\Bigr]\d\omega \\
     & & \qquad=\dfrac{mga\Xi+ge\Bigl(Qr_H-G(a+l)(\Xi-1)\Bigr)+g^2\Bigl(Gr_H+Q(a+l)(\Xi-1)\Bigr)}{2\pi\Bigl(r_H^2+(a+l)^2\Bigr)}; \label{Hawkingdeg}
     \eea
  and the energy-momentum tensor current flux is
     \bea
     F_H & = & \int\limits_0^\infty\dfrac\omega{2\pi}\Bigl[N_{m,e,g}(\omega)-N_{-m,-e,-g}(\omega)\Bigr]\d\omega \\
     & & \qquad=\dfrac1{4\pi}\left(\dfrac{ma\Xi+e\Bigl(Qr_H-G(a+l)(\Xi-1)\Bigr)+g\Bigl(Gr_H+Q(a+l)(\Xi-1)\Bigr)}{r_H^2+(a+l)^2}\right)^2+\dfrac\pi{12}T_h^2. \label{Hawkingradiation}
     \eea

  Comparing equations (\ref{fluxdem}--\ref{fluxdeg}, \ref{flux}) with equations (\ref{Hawkingdem}, \ref{Hawkingdeq}, \ref{Hawkingdeg}, \ref{Hawkingradiation}), we can conclude that the fluxes of the $U(1)$ gauge current, electric current, magnetic current and the energy-momentum tensor required to cancel gauge or gravitational anomalies at horizon are identical to that of Hawking radiation.


\section{Hawking radiation at Cosmological horizon}

  Near the cosmological horizon (CH), we must take the ingoing modes into consideration. Inside the CH $(r_c-\epsilon<r<r_c)$, the modes need to satisfy the equation $\partial_rJ^r_{(c)}=-\dfrac{\alpha\partial_r\tilde A_t}{4\pi}$, but in $r<r_c-\epsilon$ there is no anomaly, so we have a conservation equation $\partial_rJ^r_{(c-out)}=0$.
  We can get from these equation, the consistent current
     \be
     J^r=J^r_{(c-out)}\Theta(r_c-\epsilon-r)+J^r_{(c)}[1-\Theta(r_c-\epsilon-r)].
     \ee

  Under gauge transformation, the variation of the quantum effective action (\ref{acFeynman}) becomes
     \be
     -\delta\cl W=\int\d t\d r\lambda\left[\delta(r_c-\epsilon-r)\Bigl(J^r_{(c)}-J^r_{(c-out)}+\dfrac\alpha{4\pi}\tilde A_t\Bigr)+\partial_t\Bigl(\dfrac\alpha{4\pi}\tilde A_t\left[1-\Theta(r_c-\epsilon-r)\right]\Bigr)\right].
     \ee
  Because the covariant current must vanish at the CH, we obtain the $U(1)$ gauge charge flux, the electric current flux and the magnetic current flux at the CH, are determined by
     \bea
     \tilde f_m & = & \dfrac m{2\pi}\tilde A_t(r_c)=-\dfrac{m^2a\Xi+me\Bigl(Qr_c-G(a+l)(\Xi-1)\Bigr)+mg\Bigl(Gr_c+Q(a+l)(\Xi-1)\Bigr)}{2\pi\Bigl(r_c^2+(a+l)^2\Bigr)}, \label{fluxdemCH} \\
     \tilde f_e & = & \dfrac e{2\pi}\tilde A_t(r_c)=-\dfrac{mea\Xi+e^2\Bigl(Qr_c-G(a+l)(\Xi-1)\Bigr)+eg\Bigl(Gr_c+Q(a+l)(\Xi-1)\Bigr)}{2\pi\Bigl(r_c^2+(a+l)^2\Bigr)}, \label{fluxdeqCH} \\
     \tilde f_g & = & \dfrac g{2\pi}\tilde A_t(r_c)=-\dfrac{mga\Xi+ge\Bigl(Qr_c-G(a+l)(\Xi-1)\Bigr)+g^2\Bigl(Gr_c+Q(a+l)(\Xi-1)\Bigr)}{2\pi\Bigl(r_c^2+(a+l)^2\Bigr)}. \label{fluxdegCH}
     \eea

  Similarly, in $r_c-\epsilon<r<r_c$, the energy-momentum tensor satisfies the Ward identity $\partial_rT^r_{t(c)}=J^r_{(c)}\partial_r\tilde A_t+\tilde A_t\partial_rJ^r_{(c)}-\partial_rN^r_t$, and, in $r<r_c-\epsilon$, the energy-momentum satisfies $\partial_rT^r_{t(c-out)}=\tilde a_o\partial_r\tilde A_t$.
  If we use the energy-momentum tensor
    \be
    T^r_t=T^r_{t(c-out)}\Theta(r_c-\epsilon-r)+T^r_{t(c)}[1-\Theta(r_c-\epsilon-r)],
    \ee
  after general coordinate transformations, the variation of the effective action (\ref{acFeynman}) become
    \bea
    -\delta_\lambda\cl W & = & \int\d t\d r\lambda^t\left[\tilde c_o\partial_r\tilde A_t+\partial_r\Bigl(\dfrac\alpha{4\pi}\tilde A_t^2+N_t^r\Bigr)\left[1-\Theta(r_c-\epsilon-r)\right]\right. \nonumber \\
    & & \qquad+\left.\Bigl(T_{t(c)}^r-T^r_{t(c-out)}+\dfrac\alpha{4\pi}\tilde A_t^2+N_t^r\Bigr)\delta(r_c-\epsilon-r)\right];
    \eea
  and, because the covariant energy-momentum tensor must vanish at the CH, the flux of the energy-momentum tensor required to restore general coordinate covariance at quantum level in the effective field theory is
    \be
    \tilde a_o=-\dfrac1{4\pi}\left(\dfrac{ma\Xi+e\Bigl(Qr_c-G(a+l)(\Xi-1)\Bigr)+g\Bigl(Gr_c+Q(a+l)(\Xi-1)\Bigr)}{r_c^2+(a+l)^2}\right)^2-\dfrac\pi{12}T_c^2;
    \label{fluxCH}
    \ee
  where $T_c=-\dfrac1{4\pi}\dfrac{\cl C}{R^2}\Bigl\vert_{r=r_c}$ is the temperature of black hole.

  The Hawking radiation spectrum of the Pleba\'nski-Demia\'nski black hole at CH is given by the Fermi-Dirac distribution
     \be
     N_{\pm m,\pm e,\pm g}=-\dfrac1{exp\Bigl(\dfrac{\omega\mp m\tilde A_{t_1}\mp e\tilde A_{t_2}\mp g\tilde A_{t_3}}{T_h}\Bigr)+1}.
     \ee

  From this distribution, the $U(1)$ gauge current flux, the electric current flux and the magnetic current flux can be obtained as
     \bea
     F_m & = & m\int\limits_0^\infty\dfrac1{2\pi}\Bigl[N_{m,e,g}(\omega)-N_{-m,-e,-g}(\omega)\Bigr]\d\omega \\
     & & \qquad=-\dfrac{m^2a\Xi+me\Bigl(Qr_H-G(a+l)(\Xi-1)\Bigr)+mg\Bigl(Gr_H+Q(a+l)(\Xi-1)\Bigr)}{2\pi\Bigl(r_H^2+(a+l)^2\Bigr)}, \label{HawkingdemCH} \\
     F_e & = & e\int\limits_0^\infty\dfrac1{2\pi}\Bigl[N_{m,e,g}(\omega)-N_{-m,-e,-g}(\omega)\Bigr]\d\omega \\
     & & \qquad=-\dfrac{mea\Xi+e^2\Bigl(Qr_H-G(a+l)(\Xi-1)\Bigr)+eg\Bigl(Gr_H+Q(a+l)(\Xi-1)\Bigr)}{2\pi\Bigl(r_H^2+(a+l)^2\Bigr)}, \label{HawkingdeqCH}
     \eea
     \bea
     F_g & = & g\int\limits_0^\infty\dfrac1{2\pi}\Bigl[N_{m,e,g}(\omega)-N_{-m,-e,-g}(\omega)\Bigr]\d\omega \\
     & & \qquad=-\dfrac{mga\Xi+ge\Bigl(Qr_H-G(a+l)(\Xi-1)\Bigr)+g^2\Bigl(Gr_H+Q(a+l)(\Xi-1)\Bigr)}{2\pi\Bigl(r_H^2+(a+l)^2\Bigr)}, \label{HawkingdegCH}
     \eea
  and the energy-momentum tensor current is
     \bea
     F_H & = & \int\limits_0^\infty\dfrac\omega{2\pi}\Bigl[N_{m,e,g}(\omega)-N_{-m,-e,-g}(\omega)\Bigr]\d\omega \\
     & & \qquad=-\dfrac1{4\pi}\left(\dfrac{ma\Xi+e\Bigl(Qr_H-G(a+l)(\Xi-1)\Bigr)+g\Bigl(Gr_H+Q(a+l)(\Xi-1)\Bigr)}{r_H^2+(a+l)^2}\right)^2-\dfrac\pi{12}T_h^2. \label{HawkingradiationCH}
     \eea

  Comparing equations (\ref{fluxdemCH}--\ref{fluxdegCH}, \ref{fluxCH}) with equations (\ref{HawkingdemCH}, \ref{HawkingdeqCH}, \ref{HawkingdegCH}, \ref{HawkingradiationCH}), we can conclude that the fluxes of the $U(1)$ gauge current, electric current, magnetic current and the energy-momentum tensor required to cancel gauge or gravitational anomalies at cosmological horizon are identical to that of Hawking radiation.


\section{Limiting cases}

  In this section, we will reduce the parameters to obtain more simple and known metrics as certain limits of PD metric (\ref{PD2}). We will first proceed with vanishing the NUT parameter $l=0$ and then we will vanish the acceleration $\alpha$ and the magnetic charge $g$. This way we obtain the Kerr-Newman-de Sitter black hole. This black hole was studied by Jiang and Wu \cite{JW-07}.
  In the second case, we will first vanish the acceleration and the magnet charge, obtaining the NUT-Kerr-Newman-de Sitter black hole. This black hole was studied in \cite{LCY-08}.
  Finally, we will test our results obtaining the de Sitter space from this two limit cases. We get the correct Hawking temperature \cite{Ji-07, APGS-08}.

\subsubsection*{Black holes without NUT parameter}

  An interesting case is obtained if there is no NUT parameter $(l=0)$. With this value, the equation (\ref{constriction}) reduces to $\dfrac{\omega^2}{a^2}k=1$. With this restriction, the parameter $\omega$ can take the value of the angular momentum $a$ and the values for the parameters $k$, $\epsilon$ and $n$ (eqs. \ref{funcmetric-e}--\ref{constriction}) are
     \be
     k=1\qquad,\qquad\qquad\epsilon=1-\alpha^2(a^2+Q^2+G^2)-\dfrac\Lambda3a^2\qquad,\qquad\qquad n=-\alpha am,
     \ee
  the coefficients $a_3$ and $a_4$ take the values
     \be\ba{l}
     \medskip a_3=2\alpha m, \\
     a_4=-\alpha^2(a^2+Q^2+G^2)-\dfrac\Lambda3a^2,
     \ea\ee
  and the metric functions (\ref{funcmetric-1}--\ref{funcmetric-5}) of the PD metric become
     \be\ba{l}
     \medskip \Omega=1-\alpha r\cos\theta, \\
     \medskip \rho^2=r^2+a^2\cos^2\theta, \\
     \medskip \Xi=1+\dfrac\Lambda3a^2, \\
     \medskip \cl P=\sin^2\theta\left[1-2\alpha m\cos\theta+\Bigl(\alpha^2(a^2+Q^2+G^2)+\dfrac\Lambda3a^2\Bigr)\cos^2\theta\right], \\
     \hat\cl Q=(r^2+a^2+Q^2+G^2-2mr)(1-\alpha^2r^2)-\dfrac\Lambda3r^2(a^2+r^2).
     \ea \label{functionswithoutnut} \ee

  This way we have obtained the accelerated Kerr-Newman-de Sitter with magnetic monopole black hole metric. If we consider $\alpha=0$ and $G=0$, we will get
     \be
     \hat\cl Q=r^2+a^2+Q^2-2mr-\dfrac\Lambda3r^2(r^2+a^2),
     \label{QsinalphaGl}
     \ee
  and the metric becomes the Kerr-Newman-de Sitter one
     \bea
     \d s^2 & = & \dfrac{r^2-2mr+a^2+Q^2-\dfrac\Lambda3r^2(r^2+a^2)}{r^2+a^2\cos^2\theta}\left(\d t-\dfrac{a\sin^2\theta}{1+\dfrac\Lambda3a^2}\d\varphi\right)^2-\dfrac{r^2+a^2\cos^2\theta}{1+\dfrac\Lambda3a^2\cos^2\theta}\d\theta^2 \nonumber \\
      & & -\dfrac{r^2+a^2\cos^2\theta}{r^2-2mr+a^2+Q^2-\dfrac\Lambda3r^2(r^2+a^2)}\d r^2-\dfrac{\sin^2\theta\left(1+\dfrac\Lambda3a^2\cos^2\theta\right)}{r^2+a^2\cos^2\theta}\left(a\d t-\dfrac{r^2+a^2}{1+\dfrac\Lambda3a^2}\d\varphi\right)^2.
     \eea
  As we assumed there is not magnetic charge, then the value of $g=0$. When we take these values in the equations (\ref{gauge-t1}--\ref{gauge-t2}), the gauge field will be as follows
     \be
     \tilde A_t(r)=-\dfrac{ma\Bigl(1+\dfrac\Lambda3a^2\Bigr)+eQr}{r^2+a^2};
     \label{potsinalphaGl}
     \ee
  and the $U(1)$ gauge charge flux, electric current flux, energy-momentum flux and Hawking temperature are
     \bea
     \medskip f_m & = & -\dfrac m{2\pi}\tilde A_t(r_H)=\dfrac{m^2a\Bigl(1+\dfrac\Lambda3a^2\Bigr)+meQr_H}{2\pi\Bigl(r_H^2+a^2\Bigr)}, \label{fluxdemsinalphaGl} \\
     \medskip f_e & = & -\dfrac e{2\pi}\tilde A_t(r_H)=\dfrac{mea\Bigl(1+\dfrac\Lambda3a^2\Bigr)+e^2Qr_H}{2\pi\Bigl(r_H^2+a^2\Bigr)}, \label{fluxdeqsinalphaGl} \\
     \medskip a_o & = & \dfrac1{4\pi}\left(\dfrac{ma\Bigl(1+\dfrac\Lambda3a^2\Bigr)+eQr_H}{r_H^2+a^2}\right)^2+\dfrac\pi{12}T_h^2, \label{fluxsinalphaGl} \\
     T_h & = & \dfrac{m(r_H^2-a^2)-r_HQ^2-\dfrac\Lambda3r_H(r_H^2+a^2)^2}{2\pi(r_H^2+a^2)^2}, \label{tempsinalphaGl}
     \eea
  respectively; these values correspond to the values obtained in \cite{JW-07}.

\subsubsection*{Non accelerated black holes}

  Another interesting case is obtained if the black hole is not accelerated $(\alpha=0)$. The equation (\ref{constriction}) takes the form $\dfrac{\omega^2k}{a^2-l^2}=1-l^2\Lambda$. In this case, for the parameters $\epsilon$ and $n$ we get
     \be
     \epsilon=1-\dfrac\Lambda3(a^2+6l^2),\qquad\qquad\qquad n=l+\dfrac\Lambda3l(a^2-4l^2),
     \ee
  while the vale $\alpha=0$ implies that the coefficients $a_3$ and $a_4$ become
     \be
     a_3=-4\dfrac\Lambda3al,\qquad\qquad\qquad a_4=-\dfrac\Lambda3a^2,
     \ee
  and the metric functions (\ref{funcmetric-1}--\ref{funcmetric-5}) of the PD metric (\ref{PD}) take the form
     \be\ba{l}
     \medskip \Omega=1, \\
     \medskip \rho^2=r^2+(l+a\cos\theta)^2, \\
     \medskip \cl P=\sin^2\theta\left(1+4\dfrac\Lambda3al\cos\theta+\dfrac\Lambda3a^2\cos^2\theta\right), \\
     \hat\cl Q=r^2+a^2+Q^2+G^2-l^2-2mr-\dfrac\Lambda3\Bigl(r^4+r^2(a^2+6l^2)+3l^2(a^2-l^2)\Bigr);
     \ea \label{functionswithoutalpha} \ee
  and we obtain the NUT-Kerr-Newman-de Sitter with magnetic monopole black hole metric. If we have no magnetic monopole, we get
     \be
     \hat\cl Q=r^2+a^2+Q^2-l^2-2mr-\dfrac\Lambda3\Bigl(r^4+r^2(a^2+6l^2)+3l^2(a^2-l^2)\Bigr),
     \label{QsinalphaG}
     \ee
  and the metric becomes the NUT-Kerr-Newman-de Sitter one
     \bea
     \d s^2 & = & \dfrac{r^2+a^2+Q^2-l^2-2mr-\dfrac\Lambda3\Bigl(r^4+r^2(a^2+6l^2)+3l^2(a^2-l^2)\Bigr)}{r^2+(l+a\cos\theta)^2}\left(\d t-\dfrac{a\sin^2\theta+4l\sin^2\frac\theta2}{1+\dfrac\Lambda3(a^2-l^2)}\d\varphi\right)^2 \nonumber \\
      & & -\dfrac{\sin^2\theta\left(1+4\dfrac\Lambda3al\cos\theta+\dfrac\Lambda3a^2\cos^2\theta\right)}{r^2+(l+a\cos\theta)^2}\left(a\d t-\dfrac{r^2+(a+l)^2}{1+\dfrac\Lambda3(a^2-l^2)}\d\varphi\right)^2-\dfrac{r^2+(l+a\cos\theta)^2}{1+4\dfrac\Lambda3al\cos\theta+\dfrac\Lambda3a^2\cos^2\theta}\d\theta^2 \nonumber \\
      & & -\dfrac{r^2+(l+a\cos\theta)^2}{r^2+a^2+Q^2-l^2-2mr-\dfrac\Lambda3\Bigl(r^4+r^2(a^2+6l^2)+3l^2(a^2-l^2)\Bigr)}\d r^2.
     \eea
  Again the value of $g=0$ because we assumed there is not magnetic charge. When we take these values in the equations (\ref{gauge-t1}--\ref{gauge-t2}), the gauge field will be as follows
     \be
     \tilde A_t(r)=-\dfrac{ma\Bigl(1+\dfrac\Lambda3(a^2-l^2)\Bigr)+eQr}{r^2+(a+l)^2};
     \label{potsinalphaG}
     \ee
  and the $U(1)$ gauge charge flux, electric current flux, energy-momentum flux and Hawking temperature are
     \bea
     \medskip f_m & = & \dfrac{m^2a\Bigl(1+\dfrac\Lambda3(a^2-l^2)\Bigr)+meQr_H}{2\pi\Bigl(r_H^2+(a+l)^2\Bigr)}, \label{fluxdemsinalphaG} \\
     \medskip f_e & = & \dfrac{mea\Bigl(1+\dfrac\Lambda3(a^2-l^2)\Bigr)+e^2Qr_H}{2\pi\Bigl(r_H^2+(a+l)^2\Bigr)}, \label{fluxdeqsinalphaG} \\
     \medskip a_o & = & \dfrac1{4\pi}\left(\dfrac{ma\Bigl(1+\dfrac\Lambda3(a^2-l^2)\Bigr)+eQr_H}{r_H^2+(a+l)^2}\right)^2+\dfrac\pi{12}T_h^2, \label{fluxsinalphaG} \\
     T_h & = & \dfrac{m\Bigl(r_H^2-(a+l)^2\Bigr)+r_H\left[2l(a+l)-Q^2-\dfrac\Lambda3\left(\Bigl[r_H^2+(a+l)^2\Bigr]^2+2l(a+l)(4l^2-a^2)\right)\right]}{2\pi\Bigl(r_H^2+(a+l)^2\Bigr)^2}, \label{tempsinalphaG}
     \eea
  respectively; these values were reported in \cite{LCY-08}. It is important to observe that expresions (\ref{QsinalphaG}--\ref{tempsinalphaG}) reduce to (\ref{QsinalphaGl}--\ref{tempsinalphaGl}) when $l=0$.

\subsubsection*{Black holes in de Sitter space}

  As we commented in the introduction, the strongest reason to use the consistent anomaly and not the covariant anomaly is because the covariant method can give wrong results \cite{APGS-08}. In order to show that our calculations are general, we can find the Hawking temperature for a black hole in de Sitter space.
  To get the de Sitter metric, we can vanish all the parameters in equation (\ref{functionswithoutnut}) or in equation (\ref{functionswithoutalpha}) except $\Lambda$. We will get in both process the final values
     \be\ba{l}
     \medskip \Omega=\Xi=1,\qquad\qquad\qquad\rho^2=R=r^2,\qquad\qquad\qquad\Theta=0, \\
     \cl P=\sin^2\theta,\qquad\qquad\qquad{\hat\cl Q}=r^2\Bigl(1-\dfrac\Lambda3r^2\Bigr).
     \ea\ee
     
  Then, the PD metric will take the form of the de Sitter metric
     \be
     \d s^2=f(r)\d t^2-\dfrac1{f(r)}\d r^2-r^2\d^2\theta-r^2\sin^2\theta\d\varphi^2,
     \ee
  where the function $f(r)$ in the metric is given by $f(r)=\dfrac{\hat\cl Q}R=1-\dfrac\Lambda3r^2$. It is trivial to see that the cosmological horizon is $r_c=\sqrt{3/\Lambda}$. When we vanish all the parameters in the gauge fields, except $\Lambda$, we find that these gauge fields take the value of zero. So, the only flux different from zero is the total energy momentum tensor flux (eq. \ref{HawkingradiationCH}) and is $a_o=-\dfrac\pi{12}T_h^2$, where the Haking temperature is
     \be
     T_h=-\dfrac1{4\pi}\partial_rf\Bigl\vert_{r=r_c}=\dfrac1{2\pi}\dfrac\Lambda3r_c=\dfrac1{2\pi}\sqrt{\dfrac\Lambda3};
     \ee
  this value was reported by \cite{Ji-07} and \cite{APGS-08}.
  This means that the equations (\ref{fluxdem}--\ref{fluxdeg}, \ref{flux}) can be used to determine the fluxes of Hawking radiation at the EH of any Petrov D-type metric while the equations (\ref{fluxdemCH}--\ref{fluxdegCH}, \ref{fluxCH}) give completely the fluxes at CH.

\section{Conclusions}

  After the Robinson-Wilczek article \cite{RW-05}, where it was shown how to get the Hawking radiation for the Schwarzschild black hole via anomalies, several groups analyzed the Hawking radiation for more general black holes.
  In this process it was necessary to make several modifications to the original method in order to get a correct expresion for the Hawking fluxes.
  In this paper, we used this analysis of Hawking radiation based on gauge and gravitational anomalies and applied it to the case of the most general D-type metric in Petrov clasification of the Einstein-Maxwell equations; that is, the so-called Pleba\'nski-Demia\'nski black-hole.
  It is important to remark that any particular combination of parameters included in the PD black hole (mass, electric and magnetic charge, angular momentum, NUT parameter, acceleration and cosmological  constant) can be obtained from our calculations and we have shown this by obtaining special cases previously reported: the Kerr-Newman-de Sitter black hole and the NUT-Kerr-Newman-de Sitter black hole. It is necessary to stress out that both black holes were obtained by reducing in two different ways but, vanishing the NUT parameter the last one reduce to the first one.
  We finally showed that, starting from any of this two metrics it is possible to get the de Sitter space and our calculations give the correct temperature Hawking for it.

  With this generalization, we are showing that near the event horizon of the PD black hole there is quantum vacuum fluctuation effect and virtual particle pairs can appear.
  The negative energy particle enter the EH (or exit the CH) and become ingoing modes (or outgoing modes); but the positive energy particle exit the EH (or enter the CH) and become outgoing modes (or ingoing modes).
  With this process the theory becomes anomalous and the anomaly can be cancelled, thus, originating Hawking radiation as a consequence.
  That is, the cancelation anomalies method works even for the most general Petrov D-type metric.

  It is necessary, of course, to find the higher spin currents to determine the complete thermal distribution of Hawking radiation and verify the existence of an underlying $W_{1+\infty}$ algebra structure that induce the thermal spectrum of the Hawking radiation.

\section*{Acknowledgements}

  I am very thankful to the Physics Department at Centro de Investigaci\'on y de Estudios Avanzados del IPN (CINVESTAV-IPN) for all the support in the elavoration of this article. I am very thankful to I.~Smoli\'c, D.~Singleton, E.C.~Vagenas, S.~F.~Psihas and especially to T.~Juri\'c for their discussions, comments and reviews. I am especially thankful to Prof. S.~Meljanac, for his comments and advises.  This work was supported by the National Counceil for Science and Technology (CONACyT--M\'exico) under the scholarship 130881 and under the project 000000000187155.


\begin{thebibliography}{199}

\bibitem{ZS-71} Ya.B.~Zeldovich, A.A.~Starobinsky, {\it Particle creation and vacuum polarization in an anisotropic gravitational field}, Zh.E.T.F. {\bf 61} (1971) 2161, [JETP {\bf 34} (1972) 1159].
\bibitem{Haw-74} S.W.~Hawking, {\it Black hole explosions?}, Nature {\bf 248} (1974) 30.
\bibitem{Haw-75} S.W.~Hawking, {\it Particle creation by black holes}, Commun. Math. Phys. {\bf 43} (1975) 199.
\bibitem{GH-77-1} G.~Gibbons, S.W.~Hawking, {\it Action integrals and partition functions in quantum gravity}, Phys. Rev. D {\bf 15} (1977) 2752.
\bibitem{GH-77-2} G.~Gibbons, S.W.~Hawking, {\it Cosmological event horizons, thermodynamics, and particle creation}, Phys. Rev. D {\bf 15} (1977) 2738.
\bibitem{SV-96} A.~Strominger, C.~Vafa, {\it Microscopic origin of the Bekenstein-Hawking entropy}, Phys. Lett. B {\bf 379} (1996) 99. [hep-th/9601029].
\bibitem{Peet-00} A.~Peet, {\it TASI lectures on black holes in string theory}. [hep-th/0008241].
\bibitem{CF-77} S.M.~Christensen, S.A.~Fulling, {\it Trace anomalies and the Hawking effect}, Phys. Rev. D {\bf 15} (1977) 2088.
\bibitem{RW-05} S.P.~Robinson, F.~Wilczek, {\it Relationship between Hawking radiation anf gravitational anomalies}, Phys. Rev. Lett. {\bf 95} (2005) 011303.
\bibitem{IUW-06-1} S.~Iso, H.~Umetsu, F.~Wilczek, {\it Hawking radiation from charged black holes via gauge and gravitational anomalies}, Phys. Rev. Lett. {\bf 96} (2006) 151302. [hep-th/0602146].
\bibitem{IUW-06-2} S.~Iso, H.~Umetsu, F.~Wilczek, {\it Anomalies, Hawking radiation and regularity in rotating black holes}, Phys. Rev. D {\bf 74} (2006) 044017. [hep-th/0606018].
\bibitem{MS-06} K.~Murata, J.~Soda, {\it Hawking radiation from rotating black holes and gravitational anomalies}, Phys. Rev. D {\bf 74} (2006) 044018. [hep-th/0606069].
\bibitem{XC-07} Z.~Xu, B.~Chen, {\it Hawking radiation from general Kerr-(anti)de Sitter black holes}, Phys. Rev. D {\bf 75} (2007) 024041. [hep-th/0612261].
\bibitem{JW-07} Q.Q.~Jiang, S.Q.~Wu, {\it Hawking radiation from rotating black holes in anti-de Sitter spaces via gauge and gravitational anomalies}, Phys. Lett. B {\bf 647} (2007) 200. [hep-th/0701002].
\bibitem{IMU-06} S.~Iso, T.~Morita, H.~Umetsu, {\it Quantum anomalies at horizon and Hawking radiations in Myers-Perry black holes}, JHEP 0704 (2007) 068. [hep-th/0612286].
\bibitem{LCY-08} K.~Lin, S.W.~Chen, S.Z.~Yang, {\it Anomalies and Hawking radiation of NUT-Kerr-Newman de Sitter black hole}, Int. J. Theor. Phys. {\bf 47} (2008) 2453.
\bibitem{PD-76} J.F.~Pleba\'nski, M.~Demia\'nski, {\it Rotating charged and uniformly accelerating mass in general relativity}, Annals Phys. {\bf 98} (1976) 98.
\bibitem{IMU-07-1} S.~Iso, T.~Morita, H.~Umetsu, {\it Higher-spin currents and thermal flux from Hawking radiation}, Phys. Rev. D {\bf 75} (2007) 124004. [hep-th/0701272].
\bibitem{IMU-07-2} S.~Iso, T.~Morita, H.~Umetsu, {\it Fluxes of higher-spin currents and Hawking radiations from charged black holes}, Phys. Rev. D {\bf 76} (2007) 064015. arXiv:0705.3494 [hep-th].
\bibitem{IMU-08-1} S.~Iso, T.~Morita, H.~Umetsu, {\it Higher-spin gauge and trace anomalies in Two-dimensional backgrounds}, Nucl. Phys. B {\bf 799} (2008) 60. arXiv:0710.0453 [hep-th].
\bibitem{IMU-08-2} S.~Iso, T.~Morita, H.~Umetsu, {\it Hawking radiation via higher-spin gauge anomalies}, Phys. Rev. D {\bf 77} (2008) 045007. arXiv:0710.0456 [hep-th].
\bibitem{BC-08} L.~Bonora, M.~Cvitan, {\it Hawking radiation $W_\infty$ algebra and trace anomalies}, JHEP 0805 (2008) 071. arXiv:0804.0198 [hep-th].
\bibitem{BCPS-08} L.~Bonora, M.~Cvitan, S.~Pallua, I.~Smoli\'c {\it Hawking fluxes, $W_\infty$ algebra and anomalies}, JHEP 0812 (2008) 021. arXiv:0808.2360 [hep-th].
\bibitem{BCPS-09} L.~Bonora, M.~Cvitan, S.~Pallua, I.~Smoli\'c {\it Hawking fluxes, fermionic currents, $W_{1+\infty}$ algebra and anomalies}, Phys. Rev. D {\bf 80} (2009) 084034. arXiv:0907.3722 [hep-th].
\bibitem{VD-06} E.C.~Vagenas, S.~Das, {\it Gravitational anomalies, Hawking radiation, and spherically symmetric black holes}, JHEP 0610 (2006) 025. [hep-th/0606077].
\bibitem{DRV-07} S.~Das, S.P.~Robinson, E.C.~Vagenas, {\it Gravitational anomalies: A recipe for Hawking radiation}, Int. J. Mod. Phys. D {\bf 17} (2008) 533. arXiv:0705,2233 [hep-th].
\bibitem{APGS-08} V.~Akhmedova, T.~Pilling, A.~de~Gill, D.~Singleton, {\it Comments on anomaly versus WKB/tunneling methods for calculating Unruh radiation}, Phys. Lett. B {\bf 673} (2009) 227. arXiv:0808.3413 [hep-th].
\bibitem{ZSV-12} A.~Zampeli, D.~Singleton, E.C.~Vagenas {\it Hawking radiation, chirality, and the principle of effective theory of gravity}, JHEP 1206 (2012) 097. arXiv:1206.0879 [gr-qc].
\bibitem{Ji-07} Q.Q.~Jiang, {\it Hawking radiation from black holes in de Sitter spaces}. Class. Quant. Grav. {\bf 24} (2007) 4391. arXiv:0705.2068 [hep-th].
\bibitem{GP-06} J.B.~Griffiths, J.~Podolsk\'y, {\it A new look at the Pleba\'nski-Demia\'nski family of solutions}, Int. J. Mod. Phys. D {\bf 15} (2006) 335.
\bibitem{Zwa-71} D.~Zwanziger, {\it Local-lagrangian field theory of electric and magnetic charges}, Phys. Rev. D {\bf 3} (1971) 880.
\bibitem{Sin-95} D.~Singleton, {\it Magnetic charge as a ''Hidden'' gauge symmetry}, Int. J. Theor. Phys. {\bf 34} (1995) 37. [hep-th/9701044].
\bibitem{Ber-96} R.~Bertlmann, {\it Anomalies in quantum field theory}, Oxford: Oxford Univ. Press, 1996.
\bibitem{Bil-08} A.~Bilal, {\it Lectures on anomalies}. arXiv:0802.0634 [hep-th].
\bibitem{AW-84} L.~Alvarez-Gaume, E.~Witten, {\it Gravitational anomalies}, Nucl. Phys. B {\bf 234} (1984) 269.
\bibitem{BK-01} R.~Bertlmann, E.~Kohlprath, {\it Two-dimensional gravitational anomalies, Schwinger terms, and dispersion relations}, Annals Phys. {\bf 288} (2001) 137. [hep-th/0011067].

\end{thebibliography}
\end{document}